\newif\ifmulticol	\multicoltrue
\newif\ifshowgit	\showgittrue		
\newif\ifgitlocal	\gitlocalfalse		
\newif\ifbiblatex	\biblatexfalse		
\newif\ifbibnum		\bibnumtrue 		
\newif\iflineno		\linenofalse
\newif\iftoc		\tocfalse

\newif\iflucida		\lucidafalse
\newif\ifcm			\cmtrue
\newif\iflibertine	\libertinefalse
\newif\ifcharter	\charterfalse

\newcommand*{\secnumdepth}{0}			
\newcommand*{\tocdepth}{2}				
\newcommand*{\reftitle}{References}		
\newcommand*{\mytitleskip}{-0.2in}		

\multicoltrue\showgittrue\gitlocaltrue\biblatexfalse\bibnumtrue

\newcommand*{\mydocfontsize}{\ifcharter11pt\else\iflibertine11pt\else10pt\fi\fi}
\newcommand*{\setcol}{\ifmulticol twocolumn\else onecolumn\fi}

\documentclass[\mydocfontsize,\setcol]{article}

\newcommand*{\pdfauthor}{Steven A.~Frank}
\newcommand*{\pdftitle}{The fundamental equations of change in statistical ensembles and biological populations}

\input natPhys.sty

\newcommand*{\cov}{\mathrm{Cov}}

\newcommand{\bdq}{\GD\mathbf{p}}
\newcommand{\bq}{\mathbf{p}}
\newcommand{\dbq}{\dd\bq}

\newcommand{\ba}{\mathbf{a}}

\newcommand{\bdz}{\GD\mathbf{z}}
\newcommand{\bz}{\mathbf{z}}

\newcommand{\zbar}{\angb{z}}

\newcommand{\fbar}{\angb{\Gf}}
\newcommand{\bdf}{\GD\bfi}
\newcommand{\bfi}{\boldsymbol{\Gf}}

\newcommand{\bF}{\mathbf{F}}
\newcommand{\bI}{\mathbf{N}}

\newcommand{\F}{\mathcal F}

\newcommand*{\GD}{\Delta}

\newcommand*{\Gf}{\phi}

\usepackage{bm}

\DeclarePairedDelimiter\abs{\lvert}{\rvert}
\DeclarePairedDelimiter\norm{\lVert}{\rVert}
\DeclarePairedDelimiter\angb{\langle}{\rangle}
\DeclarePairedDelimiter\lrb{\lbrack}{\rbrack}
\DeclarePairedDelimiter\lr{\lparen}{\rparen}
\DeclarePairedDelimiter\lrbr{\lbrace}{\rbrace}

\makeatletter
\let\oldabs\abs \def\abs{\@ifstar{\oldabs}{\oldabs*}}
\let\oldnorm\norm \def\norm{\@ifstar{\oldnorm}{\oldnorm*}}
\let\oldangb\angb \def\angb{\@ifstar{\oldangb}{\oldangb*}}
\let\oldlrb\lrb \def\lrb{\@ifstar{\oldlrb}{\oldlrb*}}
\let\oldlr\lr \def\lr{\@ifstar{\oldlr}{\oldlr*}}
\let\oldlrbr\lrbr \def\lrbr{\@ifstar{\oldlrbr}{\oldlrbr*}}
\makeatother

\newcommand*{\dd}{\textrm{d}}

\newcommand*{\Eq}[1]{eqn~\ref{eq:#1}}

\newcommand*{\ovr}[2]{{{#1}\over{#2}}}
\newcommand*{\dovr}[2]{\ovr{\dd #1}{\dd #2}}

\newcommand*{\boldrule}{\hrule height 1.2pt}
\newcommand*{\noterule}{\medskip\boldrule\medskip}	

\hypersetup{draft}

\newcommand*{\mytitle}{\pdftitle}

\newcommand*{\myauthor}{Steven A.\ Frank \& Frank J. Bruggeman}
\newcommand*{\myaddress}{SAF: Department of Ecology and Evolutionary Biology, University of California, Irvine, CA 92697--2525, USA; FJB: Systems Biology Lab, AIMMS, VU University, Amsterdam, The Netherlands}
\newcommand*{\mycontact}{\href{https://stevefrank.org}{https://stevefrank.org}; \href{http://teusinkbruggemanlab.nl/}{http://teusinkbruggemanlab.nl/}}

\setabstract{-0.07}{\iftoc-2.0\else-0.0\fi}{%
A recent article in \textit{Nature Physics} unified key results from thermodynamics, statistics, and information theory. The unification arose from a general equation for the rate of change in the information content of a system. The general equation describes the change in the moments of an observable quantity over a probability distribution. One term in the equation describes the change in the probability distribution. The other term describes the change in the observable values for a given state. We show the equivalence of this general equation for moment dynamics with the widely known Price equation from evolutionary theory, named after George Price. We introduce the Price equation from its biological roots, review a mathematically abstract form of the equation, and discuss the potential for this equation to unify diverse mathematical theories from different disciplines. The new work in \textit{Nature Physics} and many applications in biology show that this equation also provides the basis for deriving many novel theoretical results within each discipline.
}

\begin{document}

\mymaketitle

\iftoc\mytoc{-24pt}{\newpage}\fi

\noindent Nicholson et al.\autocite{nicholson20timeinformation} recently published a unification of time-information uncertainty relations in thermodynamics. Their results follow from a single equation that describes the rate of change of the first moment, $\angb{A}$, of (thermodynamic) random variables,
\begin{equation}\label{eq:nich1}
  \dovr{\angb{A}}{t}=-\cov\lr{A,\dot{I}}+\angb{\dovr{A}{t}},
\end{equation}
with $I_i =-\ln p_i$, as the ``surprisal'' of state $i$, which has the Shannon entropy as its mean value. Accordingly, $A_i$ is a property of state $i$. Nicholson et al.\autocite{nicholson20timeinformation} illustrate how various insights about irreversible thermodynamics and its relations with information theory follow directly from this equation of motion, unifying several aspects of thermodynamics. We read their work with great interest and with admiration for their steps toward theoretical unification.

In this note we point to the importance of equation \ref{eq:nich1} in evolutionary biology, where it is known as the ``Price equation'' after George Price \autocite{price72extension,price70selection}. A recent volume of the \textit{Philosophical Transactions of the Royal Society B} collected numerous articles to celebrate the 50th anniversary of Price's first publication of his famous equation \autocite{lehtonen20fifty}.

To show the connections between Nicholson et al.'s work, the Price equation, and a potentially broader unification of various disciplines, we start with the classic evolutionary theory interpretation of the Price equation \autocite{frank12naturalb}. We then extend to a more abstract version of the Price equation, which potentially unifies many seemingly distinct results in biology, physics, probability, and statistics \autocite{frank18the-price,frank20simple}.

Evolutionary theory provides understanding of the changes in frequencies of things that replicate themselves. For simplicity, we may consider the DNA sequence of a bacterial cell as a single replicator, because the whole sequence often gets replicated when it passes to offspring. 

Replicators compete for resources and mutate, which is why they change in frequency. The frequency of a replicator increases when it makes more offspring than a less fit replicator. Here, a replicator's ``fitness'' is thought of as its performance. For example, a replicator's gene may code for a protein that influences how well the organism fits to its environment---how well its traits are adapted to the conditions of life. The Price equation relates all those processes. 

Accordingly, in equation \ref{eq:nich1}, $p_i$ refers to the frequency of a replicator $i$ that changes in time due to evolutionary processes such competition and mutation. Differential success in competition (selection) influences fitness. The random variable $A_i$ corresponds to the value of a trait associated with a replicator of type $i$. The population's average value of $A$ increases if values of $A$ covary with the tendency for its associated replicator to increase in frequency. That covariance is captured by the first term in equation \ref{eq:nich1}.

The second term of the Price equation captures the changes in trait values uncorrelated with changes in replicator frequency. For example, if the environment improves, then the measured value for the trait expressed by all individuals may increase. That increase has nothing to do with differential success between replicators and so lands in the second term. Often, one can think of the second term as a change in the frame of reference that alters measurements, and the first term as direct forces that change the frequency of particular measurements \autocite{frank18the-price}.

Because of the simple way that the Price equation relates change in trait values to change in replicator frequencies, it makes sense that this equation has been recognized as a foundational expression of evolutionary theory \autocite{walsh18evolution,lehtonen20fifty}. 

Returning to Nicholson et al., they use \Eq{nich1} to derive a variety of interesting and general results about the change in the observable quantities, $A$, in thermodynamics, information, and statistics. For example, in their section \textit{Information fluctuations and intrinsic speed}, they show that the time evolution of the distance moved by a probability distribution over states is the Fisher information metric.

A similar result follows in evolutionary biology from the Price equation (\Eq{nich1}), which illustrates that the Fisher information path length between probability distributions describes the evolutionary change in biological populations\autocite{frank09natural}. That prior result also noted the close relation between the Fisher information metric, which arose from the Price equation, and the foundations of information geometry \autocite{amari09divergence}.

To illustrate the connections between the Price equation and a wide variety of fundamental equations in different scientific disciplines, we consider an abstract version of the Price equation \autocite{frank12naturalb,frank18the-price}. We use $z$ for measured values, instead of Nicholson et al.'s $a$, to avoid a notational clash with our earlier publications, which used $a$ for a different variable. We closely follow the derivation given in Frank \autocite{frank20simple}.

The equation describes the change in the average value of some property between two populations. A population is a set of things. Each thing has a property indexed by $i$. Those things with a common property index comprise a fraction, $p_i$, of the population and have average value, $z_i$, for whatever we choose to measure by $z$.

Write $\bq$ and $\bz$ as the vectors over all $i$. The population average value is $\zbar=\bq\cdot\bz=\sum p_iz_i$, summed over $i$.

A second population has matching vectors $\bq'$ and $\bz'$. Here, $p_i'$ is the fraction of the second population derived from entities with index $i$ in the first population. Similarly, $z_i'$ is the average value in the second population of members derived from entities with index $i$ in the first population. This notation expresses an abstract mapping relation between two sets, in which the common dynamical notion of frequency change and value change over time is a special case.

Let $\GD$ be the difference between the derived population and the original population, $\GD\bq=\bq'-\bq$ and $\GD\bz=\bz'-\bz$. The difference in the averages is $\GD\zbar=\bq'\cdot\bz'-\bq\cdot\bz$. By using the definitions for $\GD\bq$ and $\GD\bz$, we can write the change in the average as the abstract form of the Price equation
\begin{equation}\label{eq:price}
  \GD\zbar=\bdq\cdot\bz+\bq'\cdot\bdz.
\end{equation}
The first term, $\bdq\cdot\bz$, is the partial difference of $\bq$ holding $\bz$ constant. The second term, $\bq'\cdot\bdz$, is the partial difference of $\bz$ holding $\bq$ constant. In the second term, we use $\bq'$ as the constant value because, with discrete differences, one of the partial change terms must be evaluated in the context of the second set.

The dot product is mathematically equivalent to the statistical covariance function, $\bdq\cdot\bz=\cov(\bfi,\bz)$, in which we define $\Gf_i=\GD p_i/p_i$. We abuse standard notation to write $\bfi=\bdq/\bq$, defining the division of vectors as elementwise division of elements. 

The term $\bfi$ is Fisher's average excess in fitness in evolutionary theory, a common measure of biological fitness. In our prior work, $\bfi\equiv\ba$, clashing with the notation of Nicholson et al. Note that in the limit of small changes, $\bfi\rightarrow\dd \bq/\bq=\dd\ln\bq=-\dot{I}$, linking Nicholson et al.'s notation to ours. After matching differences to differentials with respect to time, $\GD\equiv\dd/\dd t$, \Eq{nich1} follows, noting in \Eq{price} that the second term in the differential limit is $\bq'\cdot\bdz\rightarrow\bq\cdot\dd\bz/\dd t=\angb{\dd z/\dd t}$.

Because $\bz$ is any arbitrary value associated with entities, we can take $\bz\equiv\bfi$, yielding
\begin{equation}\label{eq:canonical}
  \GD\fbar=\bdq\cdot\bfi+\bq'\cdot\bdf=0.
\end{equation}
Importantly, $\fbar$ is always zero, because $\fbar=\sum_i\GD p_i$, the total probability change, which is zero. Because we always standardize total probability to be one, we may say that \Eq{canonical} expresses the conservation law for total probability.

The conservation of total probability may seem at first glance to be a trivial standardization. But in fact that conservation is what gives common form to seemingly different disciplines. This particular conservation associates with a shift-invariance for probability values, a symmetry that constrains geometric properties for the dynamics of probability distributions and shapes the common forms of observed probability distributions \autocite{frank11a-simple,frank16common,frank18the-price,frank20simple}.

The conserved system in \Eq{canonical} leads to a variety of deep results and connections between disciplines \autocite{frank18the-price,frank20simple}. We can always proceed to nonconserved systems by the transformation $\bfi\mapsto\bz$, in which $\zbar$ is not necessarily conserved.

Given the limited space here, we confine ourselves to one example that links the Price equation, biological fitness, information theory, and a generalized notion of d'Alembert's separation between direct and inertial forces \autocite{lanczos86the-variational}.

To start, note that the most fundamental measure of biological force concerns fitness, the relative success of replicators in competition with other replicators. That force drives evolutionary change through natural selection. Working with differentials to match Nicholson et al., we noted above that Fisher's average excess in fitness is equivalent to minus the differential of the surprisal, $\bfi\rightarrow\dd \bq/\bq=\dd\ln\bq=-\dot{I}$. This observation shows the essential equivalence in the mathematics of biological fitness and the mathematics of information theory and, as in Nicholson et al., the close association between the dynamics of population statistics and information.

For differentials, the first term of the Price equation with $\bz\equiv\bfi$ is 
\begin{equation*}
  \dd\bq\cdot\bfi=\norm{\frac{\dd\bq}{\sqrt{\bq}}}^2=\F,
\end{equation*}
in which $\bfi$ is an abstract notion of direct force acting on frequency, $\norm{\cdot}$ is the Euclidean length of a vector, and $\F$ is an abstract, nondimensional expression of the Fisher information distance metric.

To match inertial forces, $\bI$, to the second Price equation term, $\bq\cdot\dd\bfi$, we define $\bI=\dd\ln\bfi$, and thus
\begin{equation*}
  \bq\cdot\dd\bfi=\dd\bq\cdot\bI.
\end{equation*}
We can also write
\begin{equation*}
  \bI = \dd\ln\bfi=\dd\ln(\dd\ln\bq)=\dd\ln^2\bq.
\end{equation*}
The relative differential, $\dd\ln$, describes relative change. The second relative differential, $\bI=\dd\ln^2\bq$, describes the relative acceleration in frequency changes. Thus, the inertial forces acting on the frame of reference can be related to an acceleration.

Combining the expressions for the direct and inertial forces yields an alternative expression for the special Price equation form of \Eq{canonical} as 
\begin{equation*}
  \lr{\bF+\bI}\cdot\dd\bq=\lr{\dd\ln\bq+\dd\ln^2\bq}\cdot\dd\bq=0.
\end{equation*}
This expression is an abstract, nondimensional generalization of d'Alembert's principle for probability distributions that conserve total probability \autocite{frank18the-price,frank17universal,frank15dalemberts}. 

D'Alembert's principle of physical mechanics generalizes Newton's second law, force equals mass times acceleration \autocite{lanczos86the-variational}. In one dimension, Newton's law is $F=-mN$, for force, $F$, and mass, $m$, times acceleration, $-N$. In our abstract nondimensional expressions, $m$ drops out, so that $F+N=0$. 

D'Alembert generalizes Newton's law to a statement about motion in multiple dimensions such that, in conservative systems, the total work for a displacement, $\dbq$, and total forces, $\bF+\bI$, is zero. Work is the distance moved multiplied by the force acting in the direction of motion. Probability distributions are conservative in total probability. Extensions to nonconservative systems arise in some cases.

Many connections arise between this abstract expression of direct and inertial forces and fundamental equations in biology, information, thermodynamics, probability, and statistics \autocite{frank18the-price}.

Finally, recent work has extended the general form of the Price equation to describe the change in population statistics that cannot be expressed as moments. The harmonic mean provides an example \autocite{frank20the-generalized}.

In summary, Nicholson et al.'s article shows how independent lines of thought from different disciplines are now converging on a common and more general understanding of change in statistical ensembles and biological populations. Their work provides the basis for a significant conceptual unification of the sciences.

\medskip\noindent\textbf{Acknowledgments:} The Donald Bren Foundation, National Science Foundation grant DEB-1939423, and DoD grant W911NF2010227 support S.A.F.'s research.  

\medskip\noindent\textbf{Author contributions:} S.A.F. and F.J.B. jointly discussed the concepts, summarized past work, and wrote the manuscript.  

\medskip\noindent\textbf{Competing interests:} The authors declare no competing interests.

\medskip\noindent\textbf{Data availability:} The authors declare that the data supporting the findings of this study are available within the paper.


\mybiblio	

\ifmulticol\begin{strip}\hbox{\null}\end{strip}\hbox{\null}\fi

\end{document}